\documentclass{amsart}

\usepackage[english]{babel}

\usepackage[utf8]{inputenc}
\usepackage[T1]{fontenc}

\usepackage{amssymb}
\usepackage{stmaryrd}
\usepackage{amsmath,amsfonts}
\usepackage[tikz]{bclogo}
\usepackage[top=4.34cm]{geometry}

\usepackage{lmodern}
\usepackage{textcomp}
\usepackage{mathtools, bm}
\usepackage{amssymb, bm}
\usepackage[unq]{unique}
\usepackage{enumerate}
\usepackage{bbm}

\usepackage{algorithm}
\usepackage{algpseudocode}
\newcommand{\Break}{\State \textbf{break}}
\newcommand{\Halt}{\State \textbf{halt}}
\newcommand{\True}{\textbf{true }}
\newcommand{\False}{\textbf{false }}

\usepackage[breaklinks]{hyperref}
\usepackage[numbers]{natbib}    

\usepackage{amsthm}
\usepackage{thmtools}
\usepackage{comment}
\usepackage{todonotes}

\newtheorem{theorem}{Theorem}[section]
\newtheorem{lemma}[theorem]{Lemma}

\newtheorem{cor}[theorem]{Corollary}

\theoremstyle{definition}

\title{A faster algorithm for Cops and Robbers}
\author{Jan Petr}
\address{Centre for Mathematical Sciences,
Wilberforce Road,
Cambridge CB3 0WA,
United Kingdom}
\email{jp895@cam.ac.uk}

\author{Julien Portier}
\address{Centre for Mathematical Sciences,
Wilberforce Road,
Cambridge CB3 0WA,
United Kingdom}
\email{jp899@cam.ac.uk}

\author{Leo Versteegen}
\address{Centre for Mathematical Sciences,
Wilberforce Road,
Cambridge CB3 0WA,
United Kingdom}
\email{lvv23@cam.ac.uk}
\date{}

\begin{document}

\maketitle
\begin{abstract}
    We present an algorithm of time complexity $O(kn^{k+2})$ deciding whether a graph $G$ on $n$ vertices is $k$-copwin. The fastest algorithm thus far had time complexity $O(n^{2k+2})$.
    
\end{abstract}

\section{Introduction}

Cops and Robbers, introduced by Nowakowski and Winkler \cite{Now} and Quillot \cite{Qui}, is a game played on a finite undirected graph $G$ between two players, one of whom controls $k$ cops while the other controls a single robber. It proceeds as follows: First, each of the cops chooses a starting vertex. Then, so does the robber. After this initialization round, the players take turns alternately, starting with the cops. During cops' turn, each cop moves to a vertex at distance at most $1$. During robber's turn, the robber moves to a vertex at distance at most $1$. At all stages of the game, multiple cops are allowed to stand on the same vertex. The cops win if at any point of the game there is a cop on the same vertex as the robber. Otherwise, that is, if the robber evades the cops forever, the robber wins.

We say that a graph $G$ is \emph{$k$-copwin} if there is a winning strategy for $k$ cops. The \emph{cop number of a graph $G$}, denoted $c(G)$, is the minimal $k$ such that $G$ is $k$-copwin. Note that the cop number is well defined, as having a cop on each vertex of the graph ensures cops' victory, hence $c(G) \leq n$.

Finding the cop number of a given graph is the central objective in the study of the game. As the cop number of a graph is the sum of cop numbers of its connected components, the attention can be restricted to connected graphs. Some connected graphs, such as Moore graphs and the incidence graph of a projective plane, have been shown to have a cop number on the order of $\sqrt{n}$, as follows from Theorem $3$ in Aigner and Fromme \cite{AigFro}. Meyniel \cite{FRANKL} conjectured that the maximum cop number among all connected graphs on $n$ vertices $c(n)$ is $O(\sqrt{n})$. This conjecture remains unresolved to this day. It was proven independently by Scott and Sudakov \cite{SCOTT} and by Lu and Peng \cite{LU} that  $c(n)=O(n2^{-(1+o(1))\sqrt{log_2(n)}})$.

The problem of determining the cop number of a given graph is EXPTIME-complete, as shown by Kinnersley \cite{exptime}. Brandt, Pettie and Uitto showed in \cite{SETH} that, conditional on the Strong Exponential Time Hypothesis, deciding whether a graph $G$ on $n$ vertices with $O(n \log^2 n)$ edges is $k$-copwin requires time $\Omega(n^{k-o(1)})$. Clarke and MacGillivray \cite{n2k+2} described an algorithm in time $O(n^{2k+2})$ that decides whether a graph $G$ is $k$-copwin. In this paper, we present an algorithm in time $O(kn^{k+2})$ for deciding whether a graph is $k$-copwin.

\section{The algorithm}

In this section we present and analyze our new algorithm for determining whether $k$ cops have a winning strategy on a graph $G$.

Let $G=(V(G),E(G))$ be a graph and $k$ a natural number. We can assume $V(G)=[n]$. For ease of presentation, we order the cops arbitrarily and think of them as moving one after another in this order rather than simultaneously.

We define the state space $$S=\{(p_0,p_1,\ldots, p_k,t) \in [n]^{k+1}\times \mathbb{Z}_{k+1}\}.$$ 

The coordinates of these state space vectors are to be interpreted as follows. The last coordinate $t$ tells us which piece has been moved last, meaning that the next piece to move is the robber if $t=k$ and the $(t+1)$-st cop if $t \neq k$. The other coordinates describe the positions of the pieces in the graph, where $p_0$ is the position of the robber and $p_i$ for $i \in [k]$ is the position of the $i$-th cop. We say that a state $s$ is \emph{cop-winning} if the cops can ensure to catch the robber from $s$.

We now define an oriented graph $H$ with $V(H)=S$. A pair $(q,s)=((p_0,\ldots,p_k,t),(p_0',\ldots,p_k',t'))$ of states belongs to the edge set $E(H)$ if and only if:

\begin{enumerate}
    \item $t'=(t+1) \mod(k+1)$, and
    \item $(p_t,p_t') \in E(G)$ or $p_t=p_t'$, and
    \item $p_i=p_i'$ for all $i \neq t$.
\end{enumerate}

An edge $(q,s)\in E(H)$ thus represents that it is possible to move from state $q$ to state $s$. 

For a state $s$, we denote by $I(s)$ its in-neighbours $\{q \in S: (q,s) \in E(H)\}$. We will also use the following data structures:

\begin{itemize}
    \item $\mathrm{COPSWIN}$: a boolean array indexed by $S$, initialized at all $0$. 

\item $\mathrm{COUNTER}$: an array of non-negative integers indexed by those states $q=(p_0,\ldots,p_k,t)$ with $t=0$. Initialized with out-degrees of the respective states, that is, with $1+\mathrm{deg}_G(p_0)$.

\item $\mathrm{QUEUE}$: a queue of states, empty at the beginning.
\end{itemize}

$\mathrm{COPSWIN}$ is intended to mark states from which cops can win, $\mathrm{COUNTER}$ counts the number of neighbours of a state to which the robber can still move without becoming surely captured.

The algorithm is described in Algorithm \ref{alg:copwin}.

\begin{algorithm}
\caption{Determining if a graph is $k$-copwin}\label{alg:copwin}

\begin{algorithmic}
 \State \textbf{Input: }$G=([n],E(G))$, $k \in \mathbb{N}$
 \State \textbf{Output: } \True  if $G$ is $k$-copwin, \False if $G$ is not $k$-copwin
\State
\If{$k \geq n$}
\State \Return \True
\Halt
\EndIf
\State
\State Initiate $\mathrm{COPSWIN}$, $\mathrm{COUNTER}$, $\mathrm{QUEUE}$
\State
\For{$s=(p_0,p_1,\ldots, p_k, t) \in S$}
    \For{$i \in [k]$}
        \If{$p_0=p_i$}
            \State enqueue $s$ in $\mathrm{QUEUE}$
            \State $\mathrm{COPSWIN}(s):=1$
            \Break
        \EndIf
    \EndFor
\EndFor
\State
\While{$\mathrm{QUEUE}$ not empty}
    \State dequeue $s=(p_0,p_1,\ldots, p_k, t)$ from QUEUE
    \If{$t \neq 0$}
        \For{$q \in I(s)$}          \If{$\mathrm{COPSWIN}(q)=0$}
                \State enqueue $q$ in $\mathrm{QUEUE}$
                \State $\mathrm{COPSWIN}(q):=1$
            \EndIf
        \EndFor
    \Else
       \For{$q \in I(s)$}
           \State $\mathrm{COUNTER}(q):=\mathrm{COUNTER}(q)-1$ \If{$\mathrm{COUNTER}(q)=0$}
                \If{$\mathrm{COPSWIN}(q)=0$}
                    \State enqueue $q$ in $\mathrm{QUEUE}$
                    \State $\mathrm{COPSWIN}(q):=1$
                \EndIf
            \EndIf
        \EndFor
    \EndIf
\EndWhile
\State
\If{there exists $(p_1,\ldots,p_k) \in [n]^k$ s.t. for all $p_0 \in [n]: \mathrm{COPSWIN}((p_0,p_1,\ldots,p_k,0))=1$}
\State \Return \True
\Else
\State \Return \False
\EndIf

\end{algorithmic}
\end{algorithm}

We begin by observing that each state is enqueued at most once, which ensures that the algorithm terminates.

\begin{lemma}\label{Queue}
Every $s \in S$ is enqueued at most once.
\end{lemma}

\begin{proof}
Only states $s$ with $\mathrm{COPSWIN}(s)=0$ get enqueued. Whenever a state $s$ is enqueued, $\mathrm{COPSWIN}(s)$ becomes $1$.
\end{proof}

We verify that cops-winning states are exactly those whose $\mathrm{COPSWIN}$ value is $1$:

\begin{lemma}
Let $s$ be a state in $S$. Once Algorithm \ref{alg:copwin} finishes, $\mathrm{COPSWIN}(s)=1$ if and only if $s$ is cop-winning.
\end{lemma}

\begin{proof}
We first show that if $\mathrm{COPSWIN}(s)=1$ then $s$ is cop-winning. To this end, assume for contradiction that $q=(p_0,\ldots,p_k,t)$ is the first state that was assigned $\mathrm{COPSWIN}(q)=1$ during the run of Algorithm \ref{alg:copwin} without being cop-winning. All states $s$ assigned $\mathrm{COPSWIN}(s)=1$ inside the first for-loop are trivially cop-winning, so $q$ had to have been assigned $\mathrm{COPSWIN}(q)=1$ inside the while-loop. Let $s=(p_0',\ldots,p_k',t')$ be the state in the while-cycle of which $q$ got enqueued. Note that this means $q \in I(s)$, and therefore $t'=(t+1) \mod (k+1)$. Also, $t=k$, as else it would be possible to move the $t$-th cop so as to move from $q$ to $s$, contradicting that $q$ is not cop-winning. The state $q$ got enqueued when $\mathrm{COUNTER}(q)$ reached zero, that is, after all its out-neighbours got enqueued. In other words, the states $(w_0,w_1,\ldots,w_k,0)$ where $w_0 \in N(p_0) \cup \{p_0\}$ and $w_i=p_i$ for all $i \in [k]$ are all cop-winning. That means that wherever the robber moves from $p_0$, the resulting state is cop-winning, a contradiction.

It remains to prove that for every cop-winning state $s$, $\mathrm{COPSWIN}(s)=1$ once Algorithm \ref{alg:copwin} finishes. Assuming this were not the case, let $q$ be a cop-winning state chosen from cop-winning states with $\mathrm{COPSWIN}(q)=0$ such that the number of moves $m$ in which cops can catch the robber no matter how the robber moves is minimal. We have $m>0$. If $t \neq k$, it is possible to move the $t$-th cop in such a way that cops can win in $m-1$ moves from the resulting state $s$. By the choice of $q$, $s$, $\mathrm{COPSWIN}(s)=1$. But then $q$ would have been enqueued in the while-cycle of $s$, a contradiction.

Finally, assume $t=k$. Every robber's move from $p_0$ to $N(p_0) \cup \{p_0\}$ is losing in fewer than $m$ moves. That means that all the states $w$ in $\{(w_0, w_1, \ldots, w_k, 0): w_0 \in {N(p_0) \cup \{p_0\}}\}=\{w \in S: (q,w) \in E(H)\}$ have been enqueued at some point, each of them decreasing $\mathrm{COUNTER}(q)$ by $1$. After the last decrement, $\mathrm{COPSWIN}(q)$ becomes $1$, a contradiction. 
\end{proof}

Denoting the average degree of a graph $G$ by $\bar{d(G)}$, we can express the time complexity of Algorithm \ref{alg:copwin} as follows.

\begin{theorem}
Algorithm \ref{alg:copwin} determines whether a graph $G$ is $k$-copwin and has time complexity $O(kn^{k+1} (\bar{d}(G)+k))$.
\end{theorem}

\begin{proof}
The correctness of the algorithm has been verified by the previous lemma.

Now we move onto the time complexity. The initiation can be done in time $O(|S|)=O(kn^{k+1})$. The first for-cycle ends up doing $O(k|S|)=O(k^2n^{k+1})$ operations.

As by Lemma \ref{Queue} each state $s\in S$ is enqueued at most once, the total number of operations inside the while-loop is $O(\sum_{s \in S} |I(s)|)=O(\sum_{s \in S} (\mathrm{deg}_G(s_0)+1))=O(kn^k\cdot\sum_{v \in V(G)} (\mathrm{deg}_G(v)+1))=O(kn^{k+1}(\bar{d}(G)+1))$.

Determining whether there exists $(p_1,\ldots,p_k) \in [n]^k$ such that $ \mathrm{COPSWIN}((p_0,p_1,\ldots,p_k,0))=1$ for all $p_0 \in [n]$ can be done in time $O(n^{k+1})$.

The time complexity of Algorithm \ref{alg:copwin} is therefore $O(kn^{k+1}) + O(k^2n^{k+1}) + O(kn^{k+1}(\bar{d}(G)+1)) + O(n^{k+1}) = O(kn^{k+1}(\bar{d}(G)+k))$.
\end{proof}

As $\bar{d}(G)<V(G)$, we have: 

\begin{cor}
Algorithm \ref{alg:copwin} has time complexity $O(kn^{k+2})$.
\end{cor}

We conclude this article by mentioning two possible adaptations of the algorithm. Firstly, the algorithm could be easily changed to determine for each state not only if it is cop-winning, but also what is the minimum number of moves required to ensure catching the robber. Its time complexity would remain $O(kn^{k+1}(\bar{d}(G)+k))$.

Secondly, the algorithm can be adapted to variants on Cops and Robbers such as the deterministic version of Zombies and Survivors, introduced by Fitzpatrick, Howell, Messinger and Pike \cite{zombies} --- the zombies take the place of cops with the difference that in each move they move along a geodesic between them and the survivor (taking place of the robber). The only adjustment to the algorithm is a different definition of $H$, and, consequently, $I(s)$ and out-degrees of states. Computing all possible neighbours that lie on a geodesic between two vertices (here, the position of a cop and of the robber) can be done in time $O(n^3)$ using Floyd-Warshall algorithm for instance. The time complexity of the algorithm would thus be $O(kn^{k+1}(\bar{d}(G)+k) + n^3)$.

\section*{Acknowledgement}

The authors would like to thank Béla Bollobás for his valuable comments.

\bibliographystyle{abbrvnat}  
\renewcommand{\bibname}{Bibliography}
\bibliography{bibliography}

\end{document}